\documentclass[%
reprint,
superscriptaddress,
nofootinbib,
nobibnotes,
amsmath,amssymb,
aps, prd
]{revtex4-2}

\usepackage{subcaption}
\usepackage{graphicx}% Include Fig. files
\usepackage{dcolumn}% Align table columns on decimal point
\usepackage{bm}% bold math
\usepackage{hyperref}% add hypertext capabilities
\usepackage{amsmath} 
\usepackage{physics}
\usepackage{amssymb}
\usepackage{soul}
\usepackage{aas_macros}% Include Fig. files
\usepackage{xcolor}

\newcommand{\fDu}[1]{\stackon[-0.3ex]{$D^{#1}$}{\kern-1.0ex\scalebox{0.7}{$\circ$}}}

\newcommand{\zD}{{\raise1.0ex\hbox{${}^{\ \circ}$}}\!\!\!\!\!D}

\usepackage{orcidlink}

\graphicspath{{./}{figures/}{figures/res_1e6_pngs/}{figures/res_1e4_pngs/}{figures/res_1e3_pngs/}{figures/res_1e2_pngs/}}

\begin{document}

\preprint{APS/123-QED}

\title{Gravitational wave signatures from the phase-transition-induced collapse of a magnetized neutron star}

\author{Anson Ka Long \surname{Yip}~\orcidlink{0009-0008-8501-3535}}
\email{kalongyip@cuhk.edu.hk}
\affiliation{Department of Physics, The Chinese University of Hong Kong, Shatin, N.T., Hong Kong}

\author{Patrick Chi-Kit \surname{Cheong}~\orcidlink{0000-0003-1449-3363}}
% \altaffiliation[]{N3AS Postdoctoral fellow}%Lines break automatically or can be forced with \\
\affiliation{Department of Physics, University of California, Berkeley, Berkeley, CA 94720, USA}
\affiliation{Center for Nonlinear Studies, Los Alamos National Laboratory, Los Alamos, NM 87545, USA}
\affiliation{Department of Physics \& Astronomy, University of New Hampshire, 9 Library Way, Durham NH 03824, USA}

\author{Tjonnie Guang Feng \surname{Li}~\orcidlink{0000-0003-4297-7365}}
%\email{tgfli@cuhk.edu.hk}
\affiliation{Institute for Theoretical Physics, KU Leuven, Celestijnenlaan 200D, B-3001 Leuven, Belgium}
\affiliation{Department of Electrical Engineering (ESAT), KU Leuven, Kasteelpark Arenberg 10, B-3001 Leuven, Belgium }

\date{\today}% It is always \today, today,

\begin{abstract}
Strong magnetic fields make neutron stars potential sources of detectable electromagnetic and gravitational-wave signals.
Hence, inferring these magnetic fields is critical to understand the emissions of neutron stars.
However, due to the lack of direct observational evidence, the interior magnetic field configuration remains ambiguous.
Here, for the first time, we show that the internal magnetic field strength along with the composition of a neutron star can be directly constrained by detecting the gravitational waves from the \emph{phase-transition-induced collapse} of a magnetized neutron star.
By dynamically simulating this collapsing event, we first find that the dominant peaks in the gravitational waveform are the fundamental $l=0$ quasi-radial $F$ mode and the fundamental $l=2$ quadrupolar $^2f$ mode.
We next show that the maximum gravitational wave amplitude $|h|_\mathrm{max}$ increases with the maximum magnetic field strength of the interior toroidal field $\mathcal{B}_\mathrm{max}$ until the maximum rest-mass density at bounce $\rho_\mathrm{max,b}$ decreases due to the increasing $\mathcal{B}_\mathrm{max}$.
We then demonstrated that the magnetic suppression of fundamental modes found in our previous work remains valid for the hybrid stars formed after the phase-transition-induced collapses.
We finally show that measuring the frequency ratio between the two fundamental modes $f_{^2f}/f_{F}$ allows one to infer $\mathcal{B}_\mathrm{max}$ and the baryonic mass fraction of matter in the mixed phase $M_\mathrm{mp} / M_{0}$ of the resulting hybrid star.
Consequently, taking $\mathcal{B}_\mathrm{max}$ and $M_\mathrm{mp} / M_{0}$ as examples, this work has demonstrated that much information inside neutron stars could be extracted similarly through measuring the oscillation modes of the stars.
% \begin{description}
% Secondary publications and information retrieval purposes.
% You may use the \texttt{description} environment to structure your abstract;
% use the optional argument of the \verb+\item+ command to give the category of each item. 
% \end{description}
\end{abstract}

\maketitle

%\tableofcontents

\section{Introduction} \label{sec:intro}
The strongest magnetic fields in the universe, up to $10^{14 - 15}$ G, are discovered on the surface of neutron stars.
Some puzzling astronomical phenomena can be explained by these highly magnetized neutron stars, including soft gamma-ray repeaters and anomalous X-ray pulsars \cite{1998Natur.393..235K,1999ApJ...510L.111H,1995ApJ...442L..17M,2000A&A...361..240M,1995A&A...299L..41V}.
Besides, it has been demonstrated that ultrahigh magnetic fields can deform neutron stars based on the field geometry.
Neutron stars become prolate by a purely toroidal field \cite{2008PhRvD..78d4045K,2009ApJ...698..541K,2012MNRAS.427.3406F}, while they become oblate by a purely poloidal field \cite{1995A&A...301..757B,2001A&A...372..594K,2012PhRvD..85d4030Y}. 
These deformations make rotating neutron stars possible sources of continuous gravitational waves \cite{1996A&A...312..675B}.

As the magnetic field governs the emissions of neutron stars, it is vital to interpret the magnetic field configurations of neutron stars.
With radio astronomical data, the dipole-spin-down model is typically used to estimate the surface magnetic field strength of neutron stars \cite{2011MNRAS.411.2471C,2011MNRAS.415.1703C}.
Recently, the Neutron Star Interior Composition Explorer (NICER) also allows for the deduction of the geometry and strength of the surface magnetic field from the X-ray emitting hotspots in pulsars \cite{2019ApJ...887L..23B,2020ApJ...889..165D}.
Nevertheless, all these measurements can only provide clues to the surface magnetic field but not the interior magnetic field of neutron stars.
Therefore, it is still challenging to determine the nature of the magnetic field inside neutron stars.

Detection of gravitational-wave signals from neutron stars provides a novel way of probing information inside the stars.
The first signal detected was produced by a binary neutron star merger GW170817 \cite{2017PhRvL.119p1101A}.
Aside from merger events, the phase-transition-induced collapse of a neutron star is also a potential scenario for producing observable gravitational wave signals.
When the rest-mass density in the neutron star core exceeds a certain threshold, a gravitational collapse is triggered by the phase transition from hadronic matter to deconfined quark matter in the core.
This collapse results in a more compact `hybrid star' composed of hadrons and deconfined quarks.
This collapse could occur in a newly born neutron star from a supernova explosion and an accreting neutron star in a binary system \cite{1999JPhG...25..195W,2009MNRAS.392...52A}. 
Dynamical simulations were employed to study the dynamics and gravitational-wave signals of such a phase-transition-induced collapse \cite{2006ApJ...639..382L,2009MNRAS.392...52A}.
These studies have demonstrated that the fundamental $l=0$ quasi-radial $F$ mode and the fundamental $l=2$ quadrupolar $^2f$ mode of the resulting hybrid star can be excited and these modes give rise to detectable gravitational wave signals.
However, the magnetic field was not considered in these studies.
Until recently, we took a magnetic field into account for the first time and investigated the properties of a magnetized hybrid star formed from phase-transition-induced collapse \cite{2024MNRAS.534.3612Y}.

In this work, we further extract the gravitational wave signatures from the \emph{phase-transition-induced collapse} of a magnetized neutron star through dynamical simulations, and demonstrate that these signatures can be used to probe the internal magnetic field strength together with the composition of the star for the first time.
We have discussed the preliminary results of this work in \cite{Yip:2023qkh}.
Specifically, we first find that the waveform is primarily composed of the fundamental $l=0$ quasi-radial $F$ mode and the fundamental $l=2$ quadrupolar $^2f$ mode.
We then show that the maximum wave amplitude $|h|_\mathrm{max}$ increases with the maximum magnetic field strength of the interior toroidal field  $\mathcal{B}_\mathrm{max}$ until the maximum rest-mass density at bounce $\rho_\mathrm{max,b}$ decreases due to the increasing $\mathcal{B}_\mathrm{max}$.
We finally demonstrate that measuring the frequency ratio between the two fundamental modes $f_{^2f}/f_{F}$ can infer $\mathcal{B}_\mathrm{max}$ and the baryonic mass fraction of matter in the mixed phase $M_\mathrm{mp} / M_{0}$.
Therefore, we use $\mathcal{B}_\mathrm{max}$ and $M_\mathrm{mp} / M_{0}$ as two examples to illustrate that analyzing the oscillation modes of neutron stars can provide us information about the interior of the stars.

\section{Numerical methods}\label{sec:num_method}
The set-ups of simulations in this work are identical to those in our previous work \cite{2024MNRAS.534.3612Y}.
Here, we briefly highlight these set-ups for completeness.
\subsection{\label{sec:models}Initial neutron star models}
Equilibrium models of neutron stars in axisymmetry are constructed by the open-sourced code \texttt{XNS} \cite{2011A&A...528A.101B,2014MNRAS.439.3541P,2015MNRAS.447.2821P,2017MNRAS.470.2469P,2020A&A...640A..44S}, which are the initial data for the simulations.
We construct these equilibrium models with a polytropic equation of state,
    \begin{equation}
    P=K \rho^\gamma,
    \end{equation}
where $P$ denotes the pressure, $\rho$ denotes the rest-mass density and we adopt a polytropic constant $K=1.6 \times 10^5$ cm$^5$ g$^{-1}$ s$^{-2}$ (which is equivalent to 110 in the unit of $c=G=M_{\odot}=1$) and a polytropic index $\gamma=2$. 
The specific internal energy $\epsilon$ on the initial time-slice is specified by
    \begin{equation}
    \epsilon=\frac{K}{\gamma-1} \rho^{\gamma-1}.
    \end{equation}
The toroidal magnetic field \emph{enclosed in the star} follows a magnetic polytropic law
    \begin{equation}
        \mathcal{B}_{\phi}=\alpha^{-1}K_{\mathrm{m}}(\rho h\varpi^2)^m,
    \end{equation}
where $\alpha$ denotes the laspe function, $K_{\mathrm{m}}$ denotes the toroidal magnetization constant, $h$ denotes the specific enthalpy, $\varpi^2=\alpha^2\psi^4r^2\sin^2\theta$, $\psi$ denotes the conformal factor, $(r,\theta)$ denote the radial and angular coordinates in 2D spherical coordinates, and $m\geq1$ denotes the toroidal magnetization index.

There are 9 models constructed in total, with `REF' corresponding to the unmagnetized reference model and the others are magnetized neutron star models.
These models are included in our previous work \cite{2022CmPhy...5..334L}.
As this work does not aim to study neutron stars with various masses, all models have a fixed baryonic mass of $M_0=1.68$ $M_\odot$, within the mass range of ordinary neutron stars.
Also, each magnetized model has the same toroidal magnetization index $m=1$ but has different values of the toroidal magnetization constant $K_{\mathrm{m}}$.
The models are sorted by increasing maximum magnetic field strength $\mathcal{B}_\mathrm{max}$, where `T1K1' has the lowest field strength, `T1K2' has the second lowest field strength, and so on.
(`T1' means the toroidal magnetization index $m=1$ and `K' represents the toroidal magnetization constant $K_{\mathrm{m}}$.)
These models allow for a phase transition within the stellar core and enable comparison with Leung et al..
An overview of the detailed properties of all nine models can be found in Table~\ref{table1}. 

\begin{table}
	\centering
	\begin{tabular}{cccccccc}
		 Model & $\rho_{\mathrm{c}}$ & $M_{\mathrm{g}}$ & $r_\mathrm{e}$ & $\mathcal{B}_{\mathrm{max}}$\\
		 & ($10^{14}$ g cm$^{-3}$) & ($M_{\odot}$) & (km) & ($10^{17}$ G)\\
		\hline
		 REF & 8.56 & 1.55  & 11.85 & 0.00\\
		 T1K1 & 8.56 & 1.55 & 11.85 & $3.45\times10^{-2}$\\
		 T1K2 & 8.56 & 1.55 & 11.85 & $6.89\times10^{-2}$\\
		 T1K3 & 8.57 & 1.55 & 11.85 & $3.44\times10^{-1}$\\
		 T1K4 & 8.63 & 1.55 & 11.92 & 1.36\\
		 T1K5 & 8.81 & 1.56 & 12.15 & 2.63\\
		 T1K6 & 9.10 & 1.58 & 14.43 & 5.52\\
		 T1K7 & 8.81 & 1.59 & 16.21 & 6.01\\
		 T1K8 & 8.27 & 1.60 & 18.64 & 6.14\\
	\end{tabular}
	\caption{\label{table1} Properties of the 9 initial neutron star models constructed by the \texttt{XNS} code \cite{2011A&A...528A.101B,2014MNRAS.439.3541P,2015MNRAS.447.2821P,2017MNRAS.470.2469P,2020A&A...640A..44S}.
	All numerical values are rounded off to two decimal places.
	$\rho_{\mathrm{c}}$ is the central rest-mass density, $M_{\mathrm{g}}$ is the gravitational mass, $r_{\mathrm{e}}$ is the equatorial radius, and $\mathcal{B}_{\mathrm{max}}$ is the maximum toroidal field strength inside the neutron star. 
    All the models have a fixed baryonic mass $M_{\mathrm{0}}=1.68$ $M_{\odot}$ and the 8 magnetized models also have the same toroidal magnetization index $m=1$.}
\end{table}

\subsection{\label{sec:evolution}Hybrid star models and evolution}
Based on the framework introduced by \cite{2006ApJ...639..382L}, we assume that the phase transition happens instantaneously in the initial time slice and is triggered by changing the original polytropic equation to a ``softer'' equation of state for describing hybrid stars.
The MIT bag model equation of state \cite{johnson1975bag} for massless and non-interacting quarks is given by
   \begin{equation}
    P_{\mathrm{q}}=\frac{1}{3}(e-4 B),
    \end{equation}
where $P_{\mathrm{q}}$ denotes the pressure of deconfined quarks, $e$ denotes the total energy density and $B$ denotes the bag constant.
We apply the ideal gas equation of state to describe the evolution of normal hadronic matter
    \begin{equation}
    P_{\mathrm{h}}=(\gamma-1) \rho \epsilon,
    \end{equation}
where $P_{\mathrm{h}}$ denotes the pressure of hadrons and $\gamma$ is chosen to be 2.

A hybrid star formed after the phase-transition-induced collapse can be made up of two or three parts: (i) a hadronic phase for the region having a rest-mass density less than the lower threshold density $\rho < \rho_\mathrm{hm}$, (ii) a mixed phase of the deconfined quarks and hadrons for the region having a rest-mass density in between the lower threshold density and the upper threshold density $\rho_\mathrm{hm} < \rho < \rho_\mathrm{qm}$, and (iii) a region of pure quark matter phase with a rest-mass density beyond $\rho > \rho_\mathrm{qm}$ (the maximum density might or might not correspond to this phase in practice).

Following \cite{2009MNRAS.392...52A}, the equation of state for hybrid stars can be expressed as follows:
    \begin{equation}
    P= \begin{cases}P_{\mathrm{h}} & \text { for } \rho<\rho_{\mathrm{hm}}, \\ \alpha_\mathrm{q} P_{\mathrm{q}}+(1-\alpha_\mathrm{q}) P_{\mathrm{h}} & \text { for } \rho_{\mathrm{hm}} \leq \rho \leq \rho_{\mathrm{qm}}, \\ P_{\mathrm{q}} & \text { for } \rho_{\mathrm{qm}}<\rho,\end{cases}
    \label{eqn6}
    \end{equation}
    where 
    \begin{equation}
    \alpha_\mathrm{q}=1-\left(\frac{\rho_{\mathrm{qm}}-\rho}{\rho_{\mathrm{qm}}-\rho_{\mathrm{hm}}}\right)^\delta
    \end{equation}
quantifies the relative contribution of hadrons and deconfined quarks to the total pressure in the mixed phase.
By using $\delta$ as the exponent, the pressure contribution due to deconfined quarks can be adjusted.
We use 3 values of $\delta \in \{1, 2, 3\}$ to vary the pressure contribution due to deconfined quarks in the mixed phase. 
We take $\rho_\mathrm{hm}= 6.97 \times 10^{14}$ g cm$^{-3}$, $\rho_\mathrm{qm}= 24.3 \times 10^{14}$ g cm$^{-3}$ and $B^{1/4}=170$ MeV. 

\subsection{\label{sec:simulations}Simulations}
Simulations are performed three times for each of the 9 equilibrium models, once for each value of $\delta \in \{1, 2, 3\}$.  
Thus, $9 \times 3 = 27$ simulations are conducted.
The stellar models are evolved in dynamical spacetime using the new general relativistic magnetohydrodynamics code \texttt{Gmunu} \cite{2020CQGra..37n5015C,2021MNRAS.508.2279C,2022ApJS..261...22C}. 
\texttt{Gmunu} adopts a multigrid method for solving the Einstein equations in the conformally flat condition approximation.

Ideal general-relativistic magnetohydrodynamics simulations are carried out in 2D cylindrical coordinates $(R,z)$.
Axisymmetry with respect to the $z$-axis and equatorial symmetry are imposed for the simulations.
The computational domain covers the region [0,100] for both $R$ and $z$ directions, with the base grid resolution $N_{R} \times N_{z} = 32 \times 32$ and allowing 6 AMR levels (effective resolution $= 1024 \times 1024$).
The refinement criteria of AMR we used are equivalent to those in \cite{2021MNRAS.508.2279C,2022CmPhy...5..334L}.
TVDLF approximate Riemann solver \cite{1996JCoPh.128...82T} , 3rd-order reconstruction method PPM \cite{1984JCoPh..54..174C} and 3rd-order accurate SSPRK3 time integrator \cite{1988JCoPh..77..439S} are used for the simulations.
The region surrounding the star is filled with an artificially low-density `atmosphere', with a rest-mass density of $\rho_\mathrm{atm} \sim 10^{-10} \rho_\mathrm{c} \left(t=0\right)$.
In addition, as the simulations are restricted to magnetized stars with a purely toroidal field in axisymmetry, no divergence cleaning method is adopted.

\subsection{\label{sec:gw_extract}Gravitational wave extraction and mode identification}
The gravitational wave signal due to the phase-transition-induced collapse is computed by the quadrupole moment formula for an axisymmetric source \cite{2016nure.book.....S} 
\begin{equation}
h=\frac{1}{2 D_{\mathrm{obs}}}\left(2 \ddot{I}_{z z}-\ddot{I}_{R R}\right),
\end{equation}
where $h$ is the gravitational wave amplitude observed in the equatorial plane, $I_{ij}$ is the quadrupole moment, $\ddot{I}_{ij}$ is its second-time derivative, and $D_{\mathrm{obs}}$ is the distance from the source. 
Since there is no unique choice for the definition of quadrupole moment in dynamical spacetime, we choose \cite{2003PhRvD..68j4020S} 
\begin{equation}\label{eqn9}
I_{i j}=\int \rho_* x^i x^j d^3 x,
\end{equation}
where $x^i$ is the spatial coordinate, $\rho_* \equiv \rho W \sqrt{\gamma}$ is the conserved rest-mass density, $\rho$ is the rest-mass density, $W \equiv 1 / \sqrt{1-v^i v_i}$ is the Lorentz factor, $v_{i}$ is the 3-velocity and $\gamma_{ij}$ is the spatial metric.
With the continuity equation
\begin{equation}
\partial_t \rho_*+\partial_i\left(\rho_* \hat{v}^i\right)=0,
\end{equation}
where $\hat{v}^i \equiv\left(\alpha v^i-\beta^i\right)$, $\alpha$ is the lapse function and $\beta^i$ is the space-like shift vector, the first time derivative of quadrupole moment $\dot{I}_{i j}$ can be computed by
\begin{equation}
\dot{I}_{i j}=\int \rho_*\left(\hat{v}^i x^j+x^i \hat{v}^j\right) d^3 x.
\end{equation}
The second time derivative $\ddot{I}_{ij}$ is then computed by the finite difference method.

After computing the waveform, we follow the method introduced by \cite{2006ApJ...639..382L} to identify the fundamental modes in the gravitational wave signal.
Specifically, we compare our results with the perturbed equilibrium models in our previous work \cite{2022CmPhy...5..334L}, which have similar structures to those of the resulting hybrid stars in our simulations (see Appendix~\ref{sec:mode_extract} for more details).

\section{\label{sec:results}Results}
\subsection{\label{sec:summary_evolve}Summary of the evolutions}
The formation dynamics and resulting properties of hybrid stars following phase-transition-induced collapse in our simulations have been discussed in detail in our previous work \cite{2024MNRAS.534.3612Y}.
Here, we briefly summarize the main features of the evolution for completeness.

The onset of quark matter in the stellar core leads to a reduction in central pressure, triggering contraction of the star.
As the core density increases and the pressure gradient balances gravity, the star undergoes damped oscillations characterized by repeated pulsations.
Each pulsation generates shock waves that dissipate kinetic energy into heat, gradually damping the oscillations and driving the system towards a new equilibrium.
The time evolution of the maximum rest-mass density $\rho_\mathrm{max}(t)$ and magnetic field strength $\mathcal{B}_\mathrm{max}(t)$ (see Fig.~1 in \cite{2024MNRAS.534.3612Y}) shows that both quantities exhibit similar damped oscillatory behavior and are coupled in phase during the formation process.

Radial profiles of the rest-mass density $\rho(r)$ and the toroidal magnetic field strength $\mathcal{B}_{\phi}(r)$ at different times (see Fig.~2 in \cite{2024MNRAS.534.3612Y}) demonstrate that the phase-transition-induced collapse causes both $\rho(r)$ and $\mathcal{B}_{\phi}(r)$ to oscillate around a new equilibrium configuration.
The resulting hybrid star eventually reaches a slightly higher central rest-mass density and maximum magnetic field strength. 
Additionally, the magnetic field inside the star becomes more concentrated towards the core, with the position of maximum magnetic field strength shifting to smaller radii.
New equilibrium profiles for $\rho(r)$ and $\mathcal{B}_{\phi}(r)$ are established at $t=10$ ms and remain unchanged at least until $t=20$ ms.

2D profiles of the rest-mass density $\rho$ and the absolute value of the toroidal magnetic field strength $|\mathcal{B}_\phi|$ at different times (see Figs.~3 and 4 in \cite{2024MNRAS.534.3612Y}) show that only about $0.1\%$ of the initial baryonic mass $M_{0}(0)$ is ejected into the surrounding atmosphere during the evolution.
The overall morphology of the 2D profiles of both $\rho$ and $|\mathcal{B}_\phi|$ within the star is well preserved throughout the process.

\subsection{\label{sec:gw_waveform}Waveform}
\begin{figure}
	\centering
    \includegraphics[width=\columnwidth, angle=0]{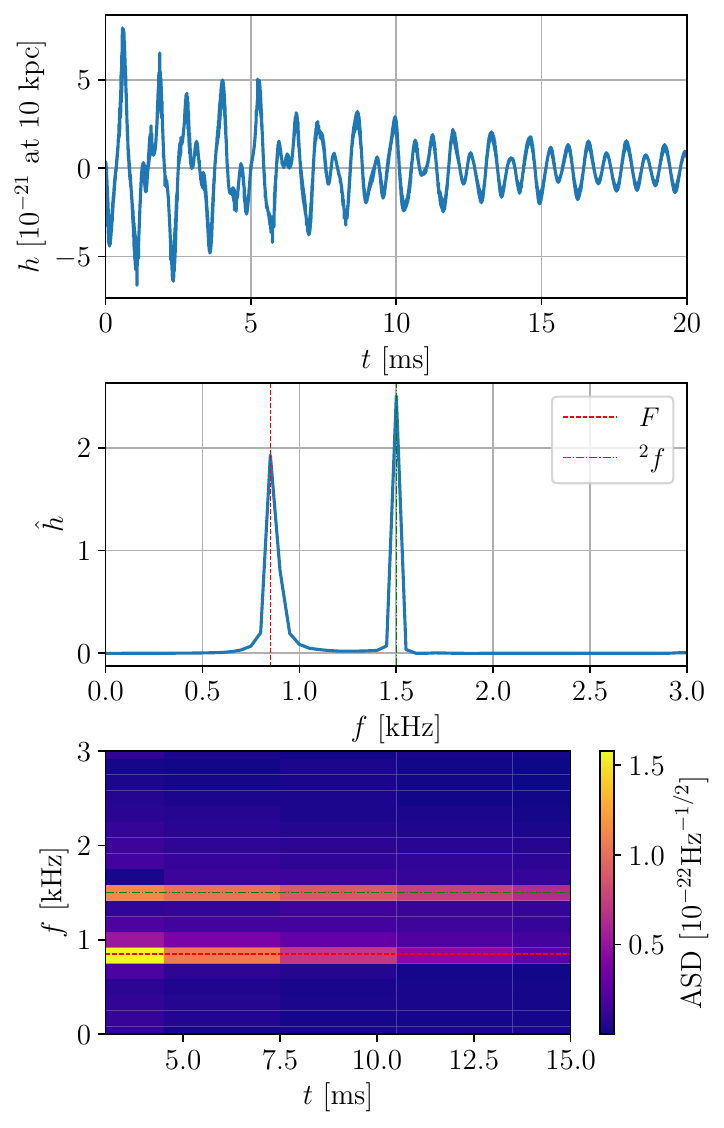}
	\caption{\label{fig1} 
            Time evolution of the gravitational wave amplitude $h$ of a magnetized hybrid star at a distance of 10 kpc (top panel), the corresponding power spectrum $\hat{h}$ in an arbitrary unit (middle panel) and the corresponding amplitude spectral density (ASD) spectrogram (bottom panel) for the simulation with the initial model T1K6 (i.e. Initial maximum magnetic field strength $\mathcal{B}_\mathrm{max} = 5.52 \times 10^{17}$ G) and an exponent quantifying the pressure contribution due to quark matter in the mixed phase $\delta=3$.
            Two dominating peaks are primarily observed in both the power spectrum and spectrogram, corresponding to the fundamental $l=0$ quasi-radial $F$ mode (red dashed line) and the fundamental $l=2$ quadrupolar $^2f$ mode (green dash-dotted line).}
\end{figure}

Similar features in the waveform are observed in different simulations, so we pick one to describe the waveform.
Here, we take the simulation with the initial magnetized neutron star model T1K6 (see Section~\ref{sec:models}) and an exponent quantifying the pressure contribution due to deconfined quarks in the mixed phase $\delta=3$ (see Section~\ref{sec:evolution}).
We plot the time evolution of the gravitational wave amplitude $h$ of a magnetized hybrid star at a distance of 10 kpc (top panel), the corresponding power spectrum $\hat{h}$ in an arbitrary unit (middle panel) and the corresponding amplitude spectral density (ASD) spectrogram (bottom panel) in Fig.~\ref{fig1}.
As mentioned in Section~\ref{sec:intro}, previous studies \cite{2006ApJ...639..382L,2009MNRAS.392...52A} have demonstrated that the fundamental $l=0$ quasi-radial $F$ mode and the fundamental $l=2$ quadrupolar $^2f$ mode of the resulting hybrid star contribute to gravitational wave signals from phase-transition-induced collapses.
Following the mode identification method introduced by \cite{2006ApJ...639..382L}, we have confirmed that our results are consistent with these established findings (see Appendix~\ref{sec:mode_extract} for more details).
The fundamental $l=0$ quasi-radial $F$ mode (red dashed line) and the fundamental $l=2$ quadrupolar $^2f$ mode (green dash-dotted line) are the 2 dominating peaks in both the power spectrum and spectrogram.

The peak of $^2f$ mode is observed in all models, while the peak of $F$ mode is observed starting from the initial model T1K4. 
The appearance of the $F$ mode peak is mainly due to the deformation of the star by a strong magnetic field. 
Similar to the effect of rotation \cite{chau1967gravitational}, the magnetic pressure breaks the spherical symmetry of the star. 
The radial mode then becomes quasi-radial and can emit gravitational waves.

\subsection{\label{sec:gw_amp}Wave amplitude}
\begin{figure}
	\centering
	\includegraphics[width=\columnwidth, angle=0]{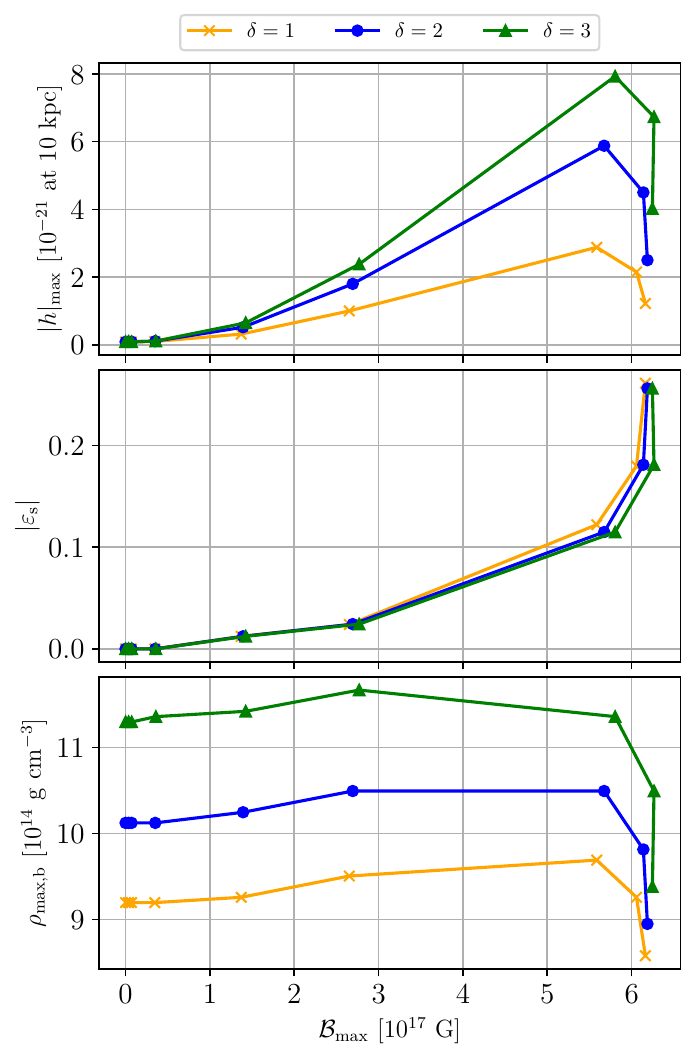}
	\caption{\label{fig2} 
             The maximum gravitational wave amplitude $|h|_\mathrm{max}$ at a distance of 10 kpc (top panel), the absolute value of the surface deformation $|\varepsilon_\mathrm{s}|$ (middle panel), and the maximum rest-mass density at bounce $\rho_\mathrm{max,b}$ (bottom panel) against the maximum magnetic field strength $\mathcal{B}_\mathrm{max}$.
             The data points are arranged into 3 sequences with 3 values of $\delta$ in \{1,2,3\}, where $\delta$ is an exponent describing the pressure contribution due to quark matter in the mixed phase.
             $|h|_\mathrm{max}$ first increases with $\mathcal{B}_\mathrm{max}$ when $\mathcal{B}_\mathrm{max} \lesssim 5 \times 10^{17} $ G.
             After reaching a maximum value at $\mathcal{B}_\mathrm{max} \sim 5 \times 10^{17} $ G, $|h|_\mathrm{max}$ drops rapidly with $\mathcal{B}_\mathrm{max}$.
             This behavior could be understood in terms of the magnetic deformation (illustrated as $|\varepsilon_\mathrm{s}|$ here) and the oscillation amplitude of the rest-mass density (described by $\rho_\mathrm{max,b}$).
             As described by Eq.~\eqref{eqn9}, the oscillation amplitude of the quadrupole moment depends on both the deformation and the rest-mass density of the star.
             In the lower $\mathcal{B}_\mathrm{max}$ regime, as $|\varepsilon_\mathrm{s}|$ increases with $\mathcal{B}_\mathrm{max}$ while $\rho_\mathrm{max,b}$ is not sensitive to $\mathcal{B}_\mathrm{max}$, which contributes to an increasing quadrupole moment and thus an increasing $|h|_\mathrm{max}$.
             On the other hand, $\rho_\mathrm{max,b}$ decreases rapidly when $\mathcal{B}_\mathrm{max} \gtrsim 5 \times 10^{17} $ so it greatly reduces the oscillation amplitude of the quadrupole moment and gives a lower $|h|_\mathrm{max}$. 
             Furthermore, increasing $\delta$ gives a larger $|h|_\mathrm{max}$, which could also result from increasing $\rho_\mathrm{max,b}$ as $\delta$ increases.
             Hence, $|h|_\mathrm{max}$ increases with $\mathcal{B}_\mathrm{max}$ until $\rho_\mathrm{max,b}$ decreases due to the increasing $\mathcal{B}_\mathrm{max}$.
	}
\end{figure}

The top panel of Fig.~\ref{fig2} shows the maximum gravitational wave amplitude $|h|_\mathrm{max}$ at a distance of 10 kpc against the maximum magnetic field strength of the resulting hybrid star $\mathcal{B}_\mathrm{max}$.
The data points are arranged into 3 sequences with 3 values of $\delta \in \{1,2,3\}$, where $\delta$ is an exponent quantifying the pressure contribution due to quark matter in the mixed phase.
First, $|h|_\mathrm{max}$ increases with $\mathcal{B}_\mathrm{max}$ when $\mathcal{B}_\mathrm{max} \lesssim 5 \times 10^{17} $ G.
After obtaining a maximum value at $\mathcal{B}_\mathrm{max} \sim 5 \times 10^{17} $ G, $|h|_\mathrm{max}$ decreases promptly with $\mathcal{B}_\mathrm{max}$.
This behavior could be interpreted in respect of magnetic deformation (illustrated as the absolute value of the surface deformation $|\varepsilon_\mathrm{s}|$ in the middle panel of Fig.~\ref{fig2}) and the oscillation amplitude of the rest-mass density (described by the maximum rest-mass density at bounce $\rho_\mathrm{max,b}$ in the bottom panel of Fig.~\ref{fig2}).
Here, we define the surface deformation as $\varepsilon_\mathrm{s}= r_\mathrm{e}/r_\mathrm{p} -1$ \cite{2012MNRAS.427.3406F}, where $r_\mathrm{p}$ and $r_\mathrm{e}$ are the polar radius and the equatorial radius of the resulting hybrid star respectively.
The detailed definitions of the polar and equatorial radius can be found in our previous work \cite{2024MNRAS.534.3612Y}. 
The maximum rest-mass density at bounce $\rho_\mathrm{max,b}$ refers to the peak value of the maximum rest-mass density during the time evolution (Similar to the central rest-mass density at bounce $\rho_\mathrm{c,b}$ in \cite{2009MNRAS.392...52A}).
According to Eq.~\eqref{eqn9}, the oscillation amplitude of the quadrupole moment depends on both the deformation and the rest-mass density of the star.
In the lower $\mathcal{B}_\mathrm{max}$ regime, as $|\varepsilon_\mathrm{s}|$ increases with $\mathcal{B}_\mathrm{max}$ while $\rho_\mathrm{max,b}$ is not sensitive to $\mathcal{B}_\mathrm{max}$, which contributes to an increasing quadrupole moment and thus an increasing $|h|_\mathrm{max}$.
On the other hand, $\rho_\mathrm{max,b}$ decreases rapidly when $\mathcal{B}_\mathrm{max} \gtrsim 5 \times 10^{17} $ so it greatly reduces the oscillation amplitude of the quadrupole moment and gives a lower $|h|_\mathrm{max}$.
Furthermore, increasing $\delta$ gives a larger $|h|_\mathrm{max}$, which could also result from increasing $\rho_\mathrm{max,b}$ as $\delta$ increases.
Accordingly, $|h|_\mathrm{max}$ increases with $\mathcal{B}_\mathrm{max}$ until $\rho_\mathrm{max,b}$ decreases due to the increasing $\mathcal{B}_\mathrm{max}$.
Besides, we also estimate the maximum value of $|h|_\mathrm{max}$ by increasing the initial baryonic mass to $M_0 = 1.80M_\odot$, which yields $|h|_\mathrm{max} \sim 10^{-20}$ (see Appendix~\ref{sec:max_amp}).

\subsection{\label{sec:gw_freq}Fundamental mode frequencies}
\begin{figure}[ht]
	\centering
	\includegraphics[width=\linewidth]{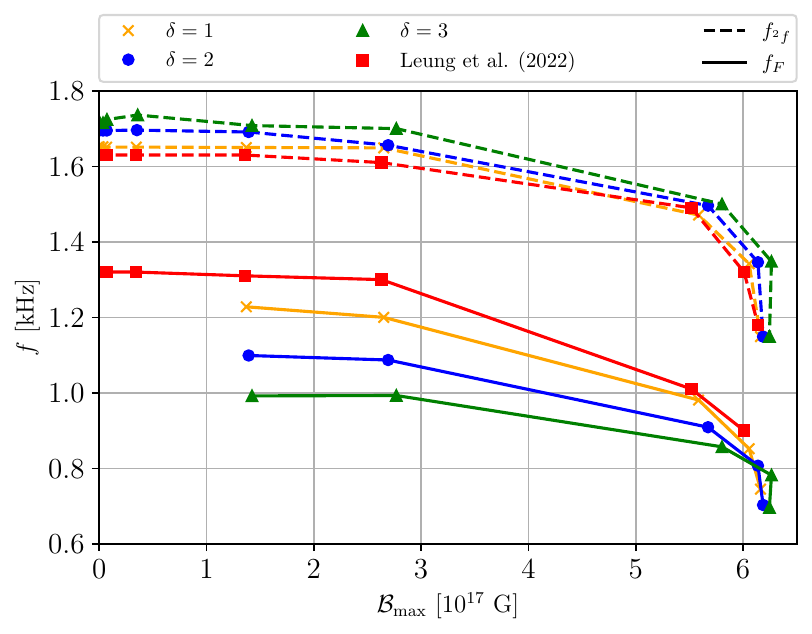}
	\caption{\label{fig3} 
             Fundamental $l=0$ quasi-radial mode frequency $f_{F}$ and fundamental $l=2$ quadrupolar mode frequency $f_{^2f}$ against the maximum magnetic field strength $\mathcal{B}_\mathrm{max}$.
             Our data points are arranged into 3 sequences according to values of $\delta$ in \{1,2,3\} used, where $\delta$ is an exponent describing the pressure contribution due to quark matter in the mixed phase.
             We also include the data points of our previous study of Leung et al. \cite{2022CmPhy...5..334L} to compare with our results.
             The fundamental modes of magnetized neutron stars without deconfined quark matter was considered in Leung et al..
             A similar decreasing trend for both fundamental modes to those in Leung et al. is observed. 
             However, $f_{F}$ in our models is smaller while $f_{^2f}$ is slightly larger than the models in Leung et al. when $\mathcal{B}_\mathrm{max} \lesssim 5 \times 10^{17} $ G.
             These frequency differences are sensitive to $\delta$.
             $f_{F}$ decreases with $\delta$ while $f_{^2f}$ increases with it. 
             Thus, the magnetic suppression of stellar oscillations found by Leung et al. is still valid in our magnetized hybrid star models and the mode frequency is also sensitive to $\delta$.}
\end{figure}

We plot fundamental $l=0$ quasi-radial mode frequency $f_{F}$ and fundamental $l=2$ quadrupolar mode frequency $f_{^2f}$ against the maximum magnetic field strength $\mathcal{B}_\mathrm{max}$ in Fig.~\ref{fig3}.
Our data points are arranged into 3 sequences according to values of $\delta \in \{1,2,3\}$ used, where $\delta$ is an exponent describing the pressure contribution due to quark matter in the mixed phase.
The data points of our previous study of Leung et al. \cite{2022CmPhy...5..334L} are also included as a comparison.
The oscillation modes of magnetized neutron stars without deconfined quark matter was considered in Leung et al..
Both fundamental modes decrease similarly to those in Leung et al.. 
Nonetheless, $f_{F}$ in our models is smaller while $f_{^2f}$ is slightly larger than the models in Leung et al. when $\mathcal{B}_\mathrm{max} \lesssim 5 \times 10^{17} $ G.
These frequency differences depend on $\delta$.
$f_{F}$ decreases with $\delta$ while $f_{^2f}$ increases with it. 
Hence, the magnetic suppression of stellar oscillations found by Leung et al. is still valid in our magnetized hybrid star models and the mode frequency is also sensitive to $\delta$.

These behaviors of $f_{F}$ and $f_{^2f}$ against $\delta$ were also observed in \cite{2009MNRAS.392...52A}.
They investigated the oscillation modes in the gravitational wave signals from the formation of unmagnetized rotating hybrid stars.
The decrease in $f_{F}$ was interpreted as the result of forming a more rapidly rotating hybrid star with a larger $\delta$, leading to a more significant mode suppression by the rotation. 
With the rediscovery of the fundamental mode behaviors in our models of magnetized non-rotating hybrid stars, these behaviors may be intrinsic properties of such a phase transition or the resulting hybrid star.
We plan to investigate this aspect thoroughly in future studies.

\subsection{\label{sec:constrain}Constraining the magnetic field and the composition}
\begin{figure}[ht]
	\centering
	\includegraphics[width=\columnwidth, angle=0]{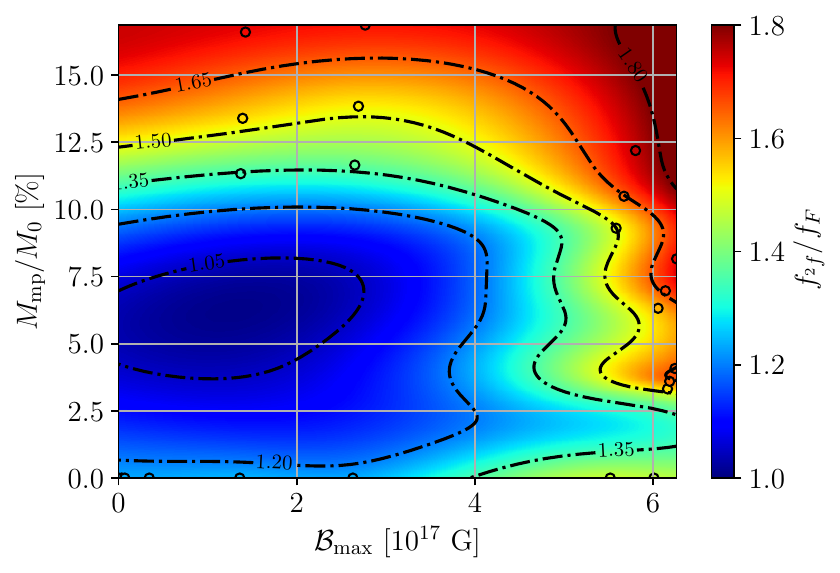}
	\caption{\label{fig4} 
             Contour plot of the frequency ratio between the fundamental $l=0$ quasi-radial mode and the fundamental $l=2$ quadrupolar mode $f_{^2f}/f_{F}$ against the maximum magnetic field strength $\mathcal{B}_\mathrm{max}$ (horizontal axis) and the baryonic mass fraction of matter in the mixed phase $M_\mathrm{mp} / M_{0}$ (vertical axis).
             This plot is constructed by the cubic radial basis function interpolation of the data points of our models and our previous work of Leung et al. \cite{2022CmPhy...5..334L}.
             The fundamental modes of magnetized neutron stars without deconfined quark matter was considered in Leung et al..
             Each data point is labeled by a colored dot with a black edge.
             The dash-dotted lines denote the contour lines for particular values of $f_{^2f}/f_{F}$.
             Values of $f_{^2f}/f_{F}$ are located in specific regions of the $\mathcal{B}_\mathrm{max}$ - $M_\mathrm{mp} / M_{0}$ plane.
             In consequence, measuring $f_{^2f}/f_{F}$ allows one to infer the value of $\mathcal{B}_\mathrm{max}$ and $M_\mathrm{mp} / M_{0}$.
	}
\end{figure}

To better illustrate the correlation between the fundamental mode frequencies and the properties of the resulting magnetized hybrid star, we plot a contour plot of the frequency ratio between the fundamental $l=0$ quasi-radial mode and the fundamental $l=2$ quadrupolar mode $f_{^2f}/f_{F}$ against the maximum magnetic field strength $\mathcal{B}_\mathrm{max}$ (horizontal axis) and the baryonic mass fraction of matter in the mixed phase $M_\mathrm{mp} / M_{0}$ (vertical axis) in Fig.~\ref{fig4}.
We constructed this plot by the cubic radial basis function interpolation of the data points of our models and in our previous work of Leung et al. \cite{2022CmPhy...5..334L}.
A colored dot with a black edge labels each data point.
The dash-dotted lines denote the contour lines for particular values of $f_{^2f}/f_{F}$. 
For a fixed value of $f_{^2f}/f_{F}$, there are localized regions in the $\mathcal{B}_\mathrm{max}$ - $M_\mathrm{mp} / M_{0}$ plane. 
Therefore, the measurement of $f_{^2f}/f_{F}$ constrains the values of $\mathcal{B}_\mathrm{max}$ and $M_\mathrm{mp} / M_{0}$.

\section{Conclusions}
In this work, for the first time, we dynamically simulate the collapse of magnetized neutron stars induced by a phase transition and demonstrate that the fundamental modes of neutron stars can be used to constrain the internal magnetic field strength together with the composition of the stars.
In particular, we first found that the waveform is primarily composed of the fundamental $l=0$ quasi-radial $F$ mode and the fundamental $l=2$ quadrupolar $^2f$ mode.
We next investigate the maximum gravitational wave amplitude $|h|_\mathrm{max}$.
$|h|_\mathrm{max}$ firstly rises with $\mathcal{B}_\mathrm{max}$ due to the increasing magnetic deformation when $\mathcal{B}_\mathrm{max} \lesssim 5 \times 10^{17} $ G. 
On the other hand, when $\mathcal{B}_\mathrm{max} \gtrsim 5 \times 10^{17} $ G, $|h|_\mathrm{max}$ decreases substantially with $\mathcal{B}_\mathrm{max}$ due to the drop in the oscillation amplitude of the rest-mass density.
Then, we have demonstrated that the magnetic suppression of stellar oscillations found in our previous work \cite{2022CmPhy...5..334L} remains valid in our models with the mode frequency value being sensitive to the pressure contribution due to quark matter $\delta$.
Finally, we have shown that the maximum magnetic field strength $\mathcal{B}_\mathrm{max}$ and the baryonic mass fraction of matter in the mixed phase $M_\mathrm{mp} / M_{0}$ could be constrained by measuring the frequency ratio between the 2 fundamental modes $f_{^2f}/f_{F}$.

The fundamental modes in this work are in the frequency range of $f \sim 600 - 1800$ Hz.
Current gravitational wave detectors, including Advanced LIGO \cite{2015CQGra..32g4001L}, Advanced Virgo \cite{2015CQGra..32b4001A}, and KAGRA \cite{2012CQGra..29l4007S,2013PhRvD..88d3007A}, are sensitive to the gravitational wave signals with frequencies of $f \sim 20 - 2000$ Hz.
Thus, these detectors can barely detect the higher frequency $^2f$ modes.
In contrast, the third-generation gravitational wave detectors, such as the Einstein Telescope (ET) \cite{2010CQGra..27s4002P} and Cosmic Explorer (CE) \cite{2017CQGra..34d4001A,2019BAAS...51c.141R,2019BAAS...51g..35R}, is designed to have a broader sensitivity band with a frequency range of $f \sim 1 - 10000$ Hz.
Moreover, it has been shown that the dominant mode frequency of the post-merger signal from binary neutron star coalescences at a distance of 68 Mpc can be measured with an accuracy of $\sim \mathcal{O}(10)$ Hz for post-merger signals with a signal-to-noise ratio of $\sim 10$ using third-generation detectors \cite{2023PhRvD.107l4009P}.
Assuming that the fundamental modes studied here can be measured with similar accuracy, and considering that the mode frequencies lie within the range $f \sim 600$–$1800$ Hz while the frequency ratio spans $f_{2f}/f_{F} \sim 1$–$1.8$ (as shown in Fig.\ref{fig4}), we expect that the frequency ratio $f_{2f}/f_{F}$ can be determined with an error of $\lesssim 3\%$.
This measurement uncertainty would translate to an error bar of comparable width ($\lesssim 3\%$) when inferring the maximum magnetic field strength $\mathcal{B}_\mathrm{max}$ and the baryonic mass fraction in the mixed phase $M_\mathrm{mp} / M_{0}$ with the contour plot in Fig.~\ref{fig4}.
However, a detailed analysis targeting the gravitational waves from phase-transition-induced collapses of neutron stars is necessary to verify this claim.
This kind of analysis is beyond the scope of this study, so we leave it for future work.

Regarding the event rate of phase-transition-induced collapses, as mentioned in Section~\ref{sec:intro}, potential hosts include both core-collapse supernovae and accreting neutron stars in binary systems. 
Given that the latter event is expected to have a much lower rate of $\sim 10^{-5}$ $\mathrm{yr}^{-1}$ in our galaxy (see e.g. \cite{2003ApJ...597.1036P}), an optimistic estimate would assume that the event rate of phase-transition-induced collapses equals that of core-collapse supernovae. 
Within our galaxy, core-collapse supernovae occur at a low rate of around 2 per 100 years \cite{2021NewA...8301498R}.
Therefore, for a reasonable event rate, it is necessary for detectors to be sensitive to sources at greater distances, such as the Virgo cluster, which is located at a distance of $\sim 15$ Mpc from Earth, where the supernova rate is expected to be $\gtrsim 1$ yr$^{-1}$ \cite{2016PhRvD..94j2001A}.

By examining $\mathcal{B}_\mathrm{max}$ and $M_\mathrm{mp} / M_{0}$ as an example, this work reveals that studying the fundamental modes of neutron stars can yield important information inside the stars.
Several extensions can be made to the current work.
Firstly, this can be done using a more realistic equation of state that considers thermal and magnetic effects.
Next, other kinds of magnetic field geometries, including purely poloidal fields and twisted torus configurations, should also be investigated.
Furthermore, since the instability of the purely toroidal field has been suppressed due to the restriction to axisymmetry in this work, 3D simulations without axisymmetry should also be performed. 
Finally, it is also necessary to consider the rotation of neutron stars, as different observations suggest that they rotate.

\begin{acknowledgments}
We acknowledge the support of the CUHK Central High-Performance Computing Cluster, on which the simulations in this work have been performed. 
This work was partially supported by grants from the Research Grants Council of Hong Kong (Project No. CUHK 14306419), the Croucher Innovation Award from the Croucher Foundation Hong Kong, and the Direct Grant for Research from the Research Committee of The Chinese University of Hong Kong. 
P.C.-K.C. acknowledges support from NSF Grant PHY-2020275 (Network for Neutrinos, Nuclear Astrophysics, and Symmetries (N3AS)).
%The data of the simulations were post-processed and visualised with 
%\texttt{yt}~\citep{2011ApJS..192....9T},
%\texttt{NumPy}~\citep{harris2020array}, 
%\texttt{pandas}~\citep{reback2020pandas, mckinney-proc-scipy-2010},
%\texttt{SciPy}~\citep{2020SciPy-NMeth}, and
%\texttt{Matplotlib}~\citep{2007CSE.....9...90H, thomas_a_caswell_2023_7697899}
\end{acknowledgments}

%\newpage
\appendix
\setcounter{figure}{0}
\renewcommand{\thefigure}{A\arabic{figure}}
\section{\label{sec:mode_extract}Mode extraction and identification}
\begin{figure}
    \centering
    \includegraphics[width=\columnwidth, angle=0]{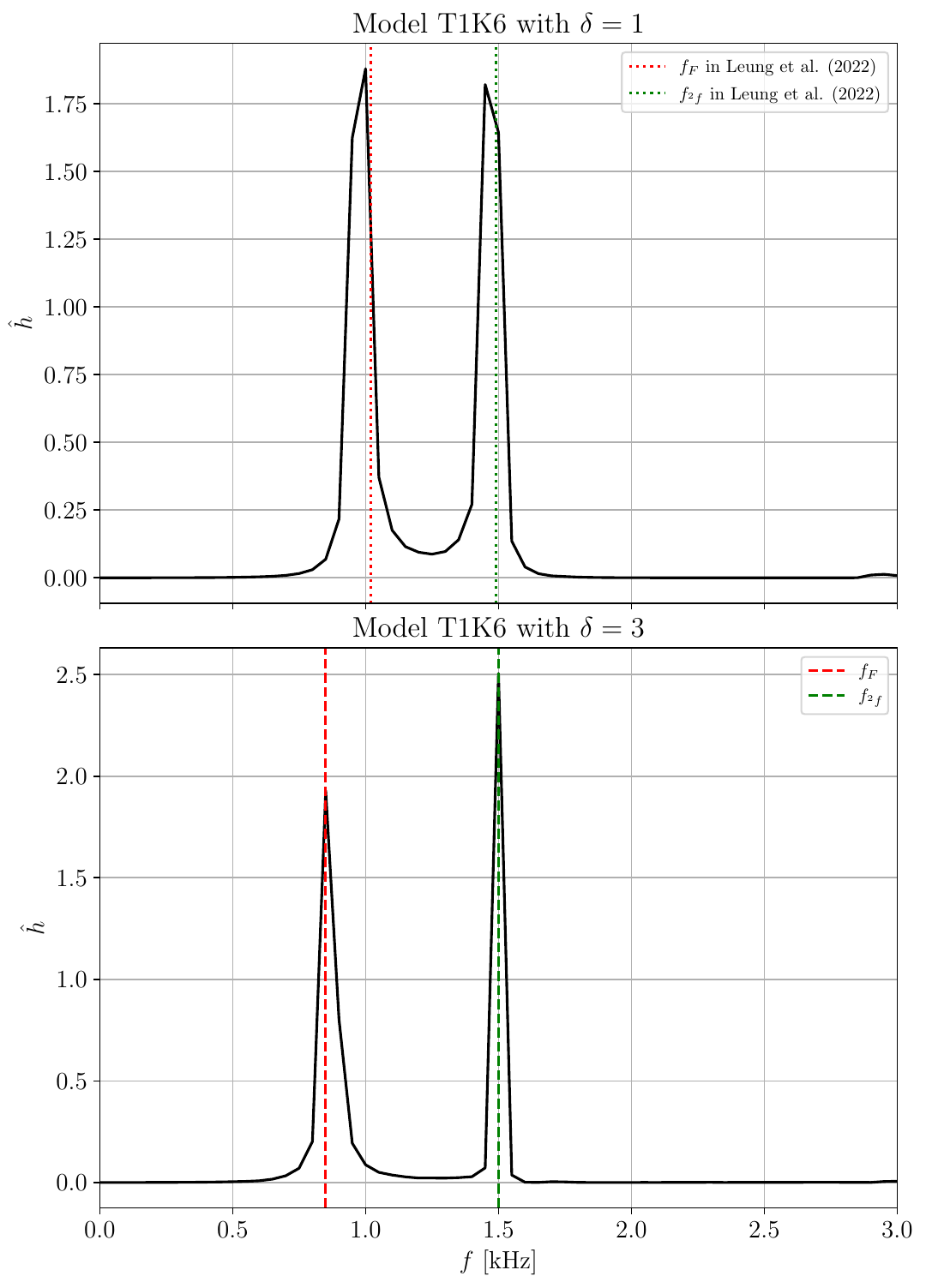}
    \caption{The gravitational wave power spectrum $\hat{h}$ (in arbitrary units) for simulations using the initial model T1K6 with pressure exponents $\delta=1$ (top panel) and $\delta=3$ (bottom panel). 
    We first identify the peaks in the power spectrum corresponding to the fundamental $l=0$ quasi-radial mode frequency $f_F$ and the fundamental $l=2$ quadrupolar mode frequency $f_{^2f}$ for the case with $\delta=1$ (top panel) by comparing them with the well-tested mode frequencies reported by Leung et al. (dotted lines). After identifying the peaks of $f_F$ and $f_{^2f}$ in the power spectrum, we track how these frequencies change as $\delta$ increases, as illustrated by the simulation with $\delta=3$ (bottom panel).}
    \label{figa1}	
\end{figure}

To illustrate our approach for identifying and extracting the fundamental modes, we present the gravitational wave power spectrum $\hat{h}$ (in arbitrary units) for simulations using the initial model T1K6 in Fig.~\ref{figa1}.
The top and bottom panels correspond to pressure exponents $\delta=1$ and $\delta=3$ respectively, which quantify the contribution from quark matter in the mixed phase.
Our identification method follows the procedure introduced by \cite{2006ApJ...639..382L}. 
Specifically, we compare the peaks in the power spectra obtained from our simulations with the well-established mode frequencies of perturbed equilibrium models reported in our previous work of Leung et al. \cite{2022CmPhy...5..334L}, which exhibit structural properties similar to those of the resulting hybrid stars in the present study.
As an illustrative example, we focus on simulations with the initial model T1K6, corresponding to the same configuration analyzed in Leung et al. \cite{2022CmPhy...5..334L}. 
As shown in the top panel of Fig.\ref{figa1}, we first identify spectral peaks corresponding to the fundamental $l=0$ quasi-radial mode frequency $f_F$ and the fundamental $l=2$ quadrupolar mode frequency $f_{^2f}$ in the $\delta=1$ case by comparing them with those reported in Leung et al.\ (dotted lines).
Once $f_F$ and $f_{^2f}$ are identified in the power spetrum, we track their evolution as the pressure exponent $\delta$ increases.
The bottom panel of Fig.~\ref{figa1} demonstrates how the peaks corresponding to $f_F$ and $f_{^2f}$ (dashed lines) are identified from the power spectrum for the simulation with $\delta=3$.

\setcounter{figure}{0}
\renewcommand{\thefigure}{B\arabic{figure}}
\section{\label{sec:max_amp}Estimating the maximum wave amplitude}
\begin{figure}[ht]
	\centering
	\includegraphics[width=\columnwidth, angle=0]{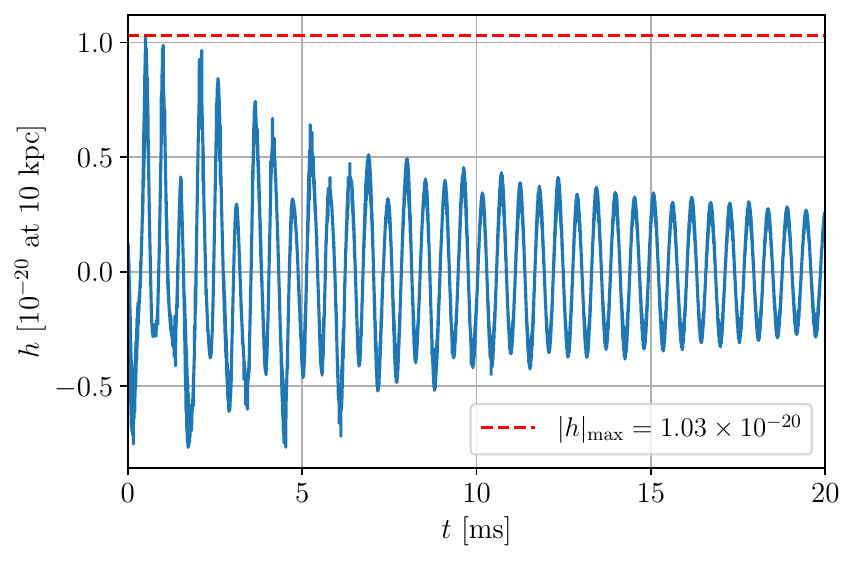}
	\caption{\label{figb1} 
             Time evolution of the gravitational wave amplitude $h$ of a magnetized hybrid star at a distance of 10 kpc, for the simulation with an initial baryonic mass $M_0 = 1.80 M_\odot$, an initial maximum magnetic field strength $\mathcal{B}_\mathrm{max} = 6.75 \times 10^{17}$ G, and an exponent quantifying the pressure contribution due to quark matter in the mixed phase $\delta = 2$. 
             A maximum gravitational wave amplitude $|h|_\mathrm{max} = 1.03 \times 10^{-20}$ (red dashed line) is observed in this simulation.
	}
\end{figure}

To estimate the maximum gravitational wave amplitude $|h|_\mathrm{max}$ that can be achieved due to phase-transition-induced collapse in the current setup, we perform a simulation with an initial neutron star whose baryonic mass is increased to $M_0 = 1.80 M_\odot$, using the same toroidal magnetization constant $K_{\mathrm{m}}$ as the T1K6 model in the main text (corresponding to a maximum toroidal magnetic field strength of $\mathcal{B}_\mathrm{max} = 6.75 \times 10^{17}$ G).
A baryonic mass of $M_0 \sim 1.80 M_\odot$ is the maximum value that can be reached with the adopted equation of state for the initial neutron star model, while maintaining the same $K_{\mathrm{m}}$ as the T1K6 model. 
This is because the T1K6 model typically yields the greatest value of $|h|_\mathrm{max}$ in all cases as shown in Section~\ref{sec:gw_amp} of the main text.
For the evolution, an exponent quantifying the pressure contribution due to quark matter in the mixed phase $\delta = 2$ is chosen, since $\delta = 3$ would cause the star to collapse into a black hole, which is beyond the scope of this study.
Fig~\ref{figb1} shows the time evolution of the gravitational wave amplitude $h$ at a distance of 10 kpc for this simulation.
We found that this simulation yields a maximum gravitational wave amplitude $|h|_\mathrm{max} = 1.03 \times 10^{-20}$ (red dashed line).
Hence, a maximum gravitational wave amplitude on the order of $10^{-20}$ can be achieved due to phase-transition-induced collapse in the current setup.

\section{\label{sec:pt_onset}Imposing a phase transition onset time}
\setcounter{figure}{0}
\renewcommand{\thefigure}{C\arabic{figure}}
To assess the stability of the magnetized neutron star models prior to the onset of the phase transition, we perform a simulation in which the phase transition is triggered a finite time after the simulation begins.
Specifically, the model is evolved for an onset time $t_\mathrm{onset} = 5$ ms before the phase transition is imposed.
Fig.~\ref{figc1} shows the time evolution of the gravitational wave amplitude $h$ at a distance of 10 kpc for the initial model T1K6 with exponent $\delta=3$, where the phase transition is triggered at $t_\mathrm{onset} = 5$ ms.
The time axis is shifted so that $t - t_\mathrm{onset} = 0$ corresponds to the onset of the phase transition, and the evolution is shown for both $t - t_\mathrm{onset} < 0$ (before the phase transition) and $t - t_\mathrm{onset} > 0$ (after the phase transition).
We observe that $h$ remains negligibly small for $t - t_\mathrm{onset} < 0$, in contrast to the significant increase in the oscillation amplitude observed for $t - t_\mathrm{onset} > 0$.
This small value of $h$ before the phase transition indicates that the quadrupole moment remains essentially unchanged, and thus the star does not undergo any significant change in its configuration prior to the onset of the phase transition.
Therefore, this demonstrates that the neutron star model remains numerically stable before the phase transition occurs.

\begin{figure}[ht]
	\centering
	\includegraphics[width=\columnwidth, angle=0]{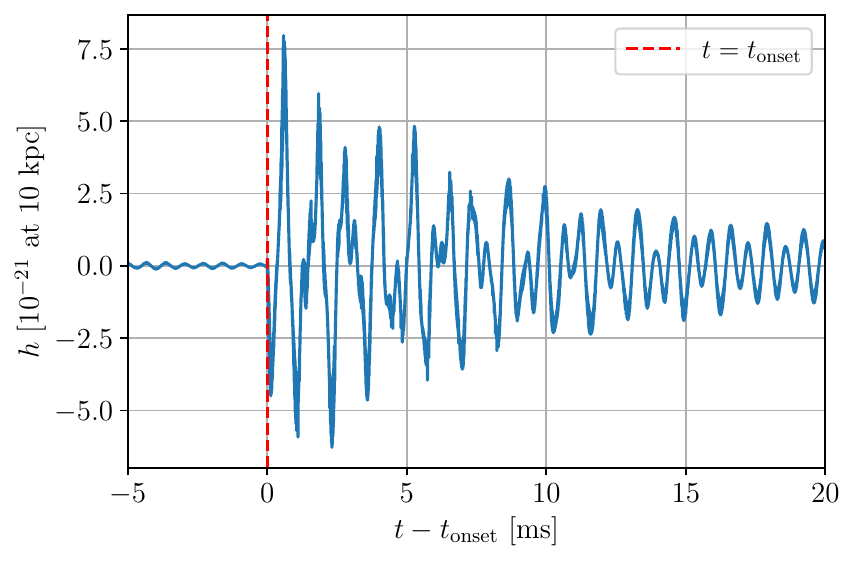}
	\caption{\label{figc1} 
             Time evolution of the gravitational wave amplitude $h$ of a magnetized hybrid star at a distance of 10 kpc, for the simulation with the initial model T1K6 and exponent $\delta=3$, with a phase transition onset time of $t_\mathrm{onset} = 5$ ms (red dashed line).
             $h$ remains negligibly small for $t - t_\mathrm{onset} < 0$, as compared to the significant increase in the oscillation amplitude observed for $t - t_\mathrm{onset} > 0$.
             This small value of $h$ prior to the phase transition illustrates that the quadrupole moment remains nearly unchanged, and thus the star does not undergo any significant change in its configuration before the onset of the phase transition.
             Hence, this shows that the neutron star model remains numerically stable before the phase transition occurs.}
\end{figure}

% \nocite{*}

% \bibliographystyle{apsrev4-2}
% \bibliographystyle{aas_marcos}
\bibliography{references}{}% Produces the bibliography via BibTeX.

\end{document}